\documentclass[1p,twocolumn,authoryear]{elsarticle}

\usepackage{amsmath}
\usepackage{amssymb}
\usepackage{graphicx}
\usepackage{tikz}

\usetikzlibrary{arrows}
\allowdisplaybreaks

\usetikzlibrary{decorations.markings}
\tikzset{
	o/.style={
		shorten >=#1,
		decoration={
			markings,
			mark={
				at position 1
				with {
					\draw[fill=black] circle [radius=#1];
				}
			}
		},
		postaction=decorate
	},
	o/.default=2pt
} 

\newcommand*{\DashedArrow}[1][]{\mathbin{\tikz [baseline=-0.25ex,-angle 60, dashed,#1] \draw [#1] (0pt,0.5ex) -- (1.3em,0.5ex);}}

\journal{Journal of Theoretical Biology}

\begin{document}
	
\tikzstyle{int}=[draw, fill=white!80, inner sep=0pt, minimum size=2.75em]
\tikzstyle{init} = [pin edge={to-,thin,black}]

\begin{frontmatter}
\date{}

\title{Periodic solutions in an SIRWS model with immune boosting and cross-immunity}

\author[UoM]{Tiffany Leung}
\author[UoM]{Barry D.~Hughes}
\author[SUT]{Federico Frascoli}
\author[UoM,MSPGH,MCRI]{James M.~McCaw\corref{cor1}}
	\ead{jamesm@unimelb.edu.au}

\cortext[cor1]{Corresponding author}

\address[UoM]{School of Mathematics and Statistics, University of Melbourne, Victoria 3010, Australia}
\address[SUT]{Department of Mathematics, Faculty of Science, Engineering and Technology, Swinburne University of Technology, Hawthorn, Victoria 3122, Australia}
\address[MSPGH]{Melbourne School of Population and Global Health, University of Melbourne, Victoria 3010, Australia}
\address[MCRI]{Modelling and Simulation, Infection and Immunity Theme, Murdoch Childrens Research Institute, Royal Children's Hospital, Parkville, Victoria 3052, Australia}

\begin{abstract}
Incidence of whooping cough, an infection caused by \textit{Bordetella pertussis} and \textit{Bordetella parapertussis}, has been on the rise since the 1980s in many countries. Immunological interactions, such as immune boosting and cross-immunity between pathogens, have been hypothesised to be important drivers of epidemiological dynamics. We present a two-pathogen model of transmission which examines how immune boosting and cross-immunity can influence the timing and severity of epidemics. We use a combination of numerical simulations and bifurcation techniques to study the dynamical properties of the system, particularly the conditions under which stable periodic solutions are present. We derive analytic expressions for the steady state of the single-pathogen model, and give a condition for the presence of periodic solutions. A key result from our two-pathogen model is that, while studies have shown that immune boosting at relatively strong levels can independently generate periodic solutions, cross-immunity allows for the presence of periodic solutions even when the level of immune boosting is weak. Asymmetric cross-immunity can produce striking increases in the incidence and period. Our study underscores the importance of developing a better understanding of the immunological interactions between pathogens in order to improve model-based interpretations of epidemiological data. 
\end{abstract}


\end{frontmatter}

\section{Introduction} \label{Introduction}

Understanding immune-mediated interactions of closely related pathogens, such as cross-immunity, can be important in explaining the epidemiological patterns of infectious diseases \citep{Adams2006, Bhattacharyya2015}. Multi-pathogen models with cross-immunity \citep{Restif2006, Restif2008, Zhang2004} are based on extensions of the susceptible-infectious-recovered-susceptible (SIRS) model of infectious disease transmission \citep{Keeling2008}, where each individual in a homogeneously mixing population is categorised into one of three classes: susceptible (S) to infection; infectious (I) if they can transmit the infection; or recovered (R) if they have cleared the infection and are (temporarily) immune. As immunity wanes over time, those who are recovered become susceptible once again.

While the mechanisms through which cross-immunity affects infection remain unclear, it is commonly assumed in mathematical models that cross-immunity acts by reducing susceptibility \citep{Kamo2002, Restif2006, Restif2008}, reducing infectivity \citep{White1998}, or by polarised immunity \citep{Gog2002, Gog2002a} in which individuals are either fully susceptible or fully immune immediately following infection. The equilibrium dynamics of multi-pathogen SIR-type models with cross-immunity have been studied to find conditions for coexistence \citep{Nuno2005, Vasco2007, White1998}, the presence of sustained oscillations \citep{Andreasen1997, Nuno2005} or the lack thereof \citep{Castillo-Chavez1989, Gog2002}. 

In a study of seroepidemiology of \textit{Bordetella pertussis} infections, \cite{Cattaneo1996} hypothesised that the maintenance of high antibody levels for pertussis components in the absence of typical pertussis symptoms may be due to immune boosting. This immunological interaction coincides with a subsequent increase (boosting) of immunity levels in individuals following re-exposure, and has been captured in mathematical models of pertussis \citep{Aguas2006, Dafilis2012, Dafilis2014a, Dafilis2014, Lavine2011}. Contrary to \cite{Aguas2006} who interpreted immune boosting as an estimate of vaccine efficacy, \cite{Lavine2011} parameterised it as the amount of antigen exposure required to stimulate a boost in immunity relative to the amount required to produce an infection in a fully susceptible (immunologically naive) individual. Furthermore, Lavine and colleagues hypothesised that the amount of antigen required to trigger a boost in immunity may be less than that required to produce a naive infection, implying that ``boosts'' may be more easily triggered than naive infections. With age structure and vaccination added to Lavine's susceptible-infectious-recovered-waning-susceptible (SIRWS) model, \cite{Lavine2011} reproduced the patterns of pertussis incidence data from Massachusetts, USA. Subsequently, \cite{Dafilis2012} demonstrated that the SIRWS model without vaccination and age structure is capable of generating damped and undamped oscillations, and chaos in the presence of seasonally forced transmission \citep{Dafilis2014a, Dafilis2014}.

Pertussis notifications have been rising since the 1980s \citep{Cherry2003}, and studies have found that infections caused by \textit{Bordetella parapertussis}---a bacteria capable of causing symptoms similar to a \textit{B.~pertussis} infection---are not uncommon \citep{Bokhari2011, Cherry2012, He1998}. The two \textit{Bordetella} pathogens are closely related, and studies using a murine model of infection have considered the level of protection gained from infection-induced immunity of one \textit{Bordetella} pathogen against subsequent infections with the other \citep{Watanabe2001a, Wolfe2007, Worthington2011}. The results from these studies are inconclusive, but suggest existence of cross-immunity and perhaps asymmetry in this interaction. The two pathogens have been observed to exhibit strikingly out of phase recurrent epidemics \citep{Lautrop1971}---behaviour that may plausibly be induced by cross-immunity. 

Our study investigates the immune-mediated interactions between two pathogens and considers how they may manifest in infectious disease epidemiology. While the application for this model is motivated by whooping cough, the same methods may be adapted to describe other multi-pathogen diseases, such as influenza \citep{Mathews2009}. In Section \ref{sec:sirws}, we introduce a two-pathogen extension of the SIRWS model that incorporates immune boosting and cross-immunity. In Section \ref{sec:methods}, we carry out the analysis to find periodic solutions in the system. The results in Section \ref{sec:results} illustrate how periodic solutions can be generated under a range of scenarios for the degree of cross-immunity and strength of immune boosting. Section \ref{sec:discussion} summarises our findings and discusses their epidemiological relevance.

\section{The SIRWS model with cross-immunity and immune boosting} \label{sec:sirws}

The SIRWS model \citep{Lavine2011} extends the SIRS model by further dividing the population of immune individuals into two classes based on their level of immunity. Those in the recovered (R) class are fully immune. Those whose immunity has waned sufficiently (W) may either lose their immunity and return to the susceptible class, or have their immunity boosted upon re-exposure and return to the recovered class. The system is mathematically represented by the following set of ordinary differential equations (ODEs) for the proportions of the population in each class:
\begin{subequations}
	\label{sirws}
	\begin{align}
	\frac{dS}{dt} &= \mu (1 - S) -\beta IS  + 2 \kappa W \, , \\
	\frac{dI}{dt} &= \beta IS - \gamma I - \mu I \, , \\
	\frac{dR}{dt} &= \gamma I - 2 \kappa R + \nu \beta IW - \mu R \, , \\
	\frac{dW}{dt} &= 2 \kappa R - 2 \kappa W - \nu \beta IW - \mu W \, ,
	\end{align}
\end{subequations}
where $\beta$ is the transmission rate, and $\gamma$ the recovery rate. Births and deaths occur at equal rates $\mu$, so that the population size remains fixed. Disease-induced mortality is not considered. Immunity is lost at a rate $\kappa$. The transition time from R to S in the absence of boosting is $2/(2\kappa + \mu)$, so that for $\mu \ll \kappa$, the average total duration of immunity is approximately $1/\kappa$. Immune boosting occurs at a rate $\nu \beta I$ proportional to the force of infection $\beta I$, where $\nu$ is the relative strength of immune boosting ($\nu \geq 0$). Allowing $\nu > 1$ implies that immune boosting may be more easily triggered than a naive infection. The addition of the immune boosting term produces limit cycles \citep{Dafilis2012}---dynamics that are qualitatively different to those of the SIRS model, a system known never to exhibit limit cycles.

\begin{figure}
	\centering
	\includegraphics[width=0.66\textwidth]{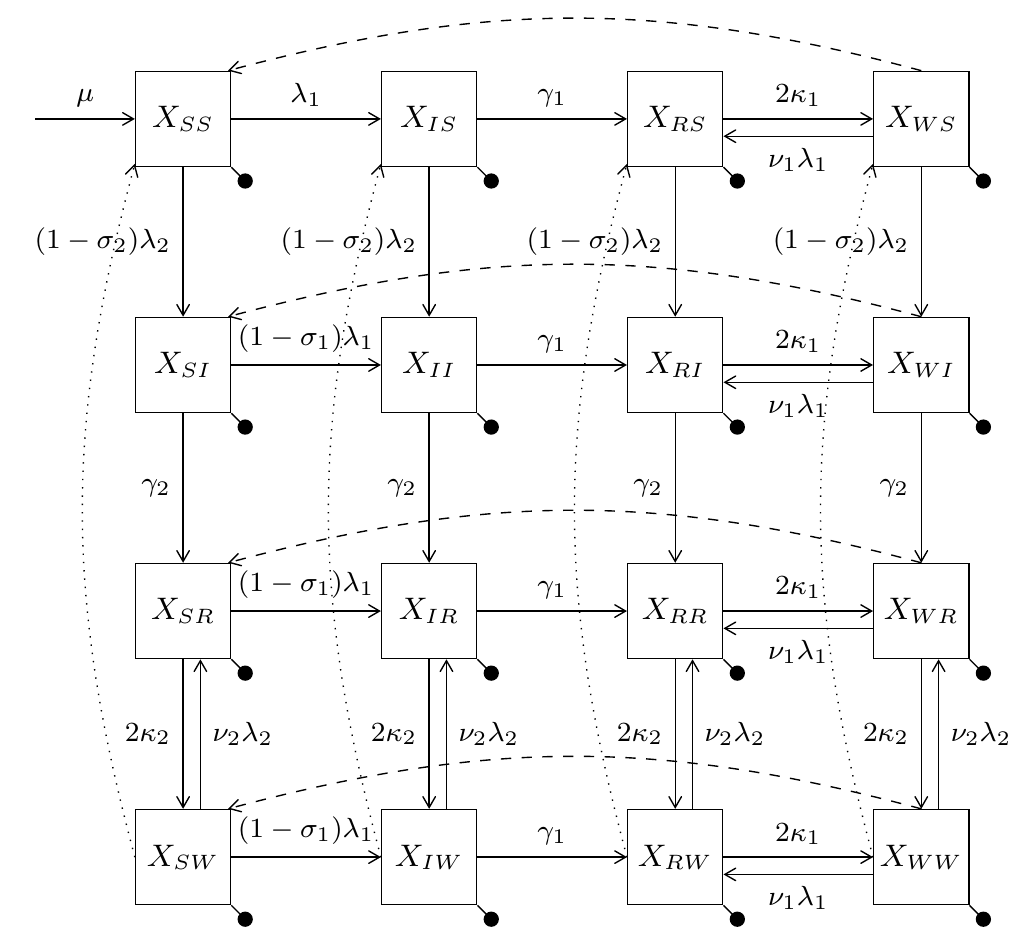}	
	\caption{A flow diagram of the two-pathogen SIRWS model. Boxes represent the compartments into which the population partitions, and arrows represent the rates at which individuals transfer between compartments. The force of infection $\lambda_i$ is the transmission coefficient times the sum of those infectious with pathogen $i = 1, 2$. For brevity, the dashed ($\DashedArrow$) and dotted ($\DashedArrow[dotted]$) arrows represent the immunity waning rates $2\kappa_1$ and $2\kappa_2$, respectively. The death rate $\mu$ is represented by a bullet (\protect\tikz \protect\fill[black] (0.4ex,0.4ex) circle (0.5ex);).}
	\label{fig:flow_diagram}
\end{figure}

Our model builds on the SIRWS framework and includes the cross-protective interactions of a second pathogen. The model is described by a system of sixteen ODEs in \ref{a:two-SIRWS} and will henceforth be referred to as the two-pathogen model. A description of the parameters is provided in Table \ref{table:params}. As depicted by a flow diagram in Figure \ref{fig:flow_diagram}, the population is divided into classes labelled $X_{mn}$, where $m$ and $n$ represent individuals with that class's disease status with respect to the first and second pathogen, for $m, n \in \{S, I, R, W\}$. Infection with one pathogen confers cross-immunity against infection with the other pathogen. Those who are infectious, recovered, or waning with respect to pathogen $j$ gain a proportional reduction in the force of infection for pathogen $i$ by a factor $(1 - \sigma_i)$, for $(i, j) = {(1, 2), (2, 1)}$. A value of $\sigma = 0$ represents no cross-immunity. As $\sigma$ approaches one, full cross-immunity is attained. Throughout the paper, we assume that cross-immunity is conferred upon infection, as opposed to recovery, and acts by reducing an individual's susceptibility to the second pathogen. However, cross-immunity may act through other ways, such as a reduction in infectivity, which may be adapted to the two-pathogen model (described in \ref{a:two-SIRWS}).

\begin{table}
	\centering
	\footnotesize{
	\begin{tabular}{p{1.2cm} p{7.5cm} p{1.6cm} p{1.6cm}}
		\hline
		Parameter & Description & Default & LHS Range \\ 
		& & & $\mathcal{U}$($a$, $b$) \\
		\hline
		$\mu$ 	 	& Birth and death rate  & 1/80 $\text{y}^{-1}$ & (0.01, 0.02) \\
		$\gamma_i$ 	& Recovery rate of infection with pathogen $i$ & 17 $\text{y}^{-1}$ & (10.4, 52) \\
		$\kappa_i$ 	& Loss of immunity rate for infection with pathogen $i$ & 1/10 $\text{y}^{-1}$ & (0.01, 1) \\
		$\nu_i$ 	& Immune boosting strength for infection with pathogen $i$ & 1 & (0, 5) \\
		$\beta_i$ 	& Rate of transmission & 260 $\text{p}^{-1} \text{y}^{-1}$ & (200, 350) \\
		$\sigma_i$ 	& Protection conferred by infection with pathogen $j$ against pathogen $i$ & [0, 1] & \\
		\hline
	\end{tabular}}
	\caption{Table of parameters and their meanings for the two-pathogen SIRWS model. All Latin hypercube sampling parameter ranges except $\beta$ were taken from \cite{Campbell2015b}. ($i = (1,2)$, y = years; p = person)}
	\label{table:params}
\end{table}

\section{Methods} \label{sec:methods}

To identify periodic solutions in the single-pathogen SIRWS model described by Equations \eqref{sirws}, we used Latin hypercube sampling (LHS) \citep{Blower1994} to simultaneously sample through all parameters from a uniform distribution using \cite{MATLAB2014}. The range of each parameter was divided into 100 equiprobable intervals. Parameter ranges, taken from a study of pertussis by \cite{Campbell2015b}, were set to an average life expectancy $1/\mu$ between 50 to 100 years; the average duration of infectiousness $1/\gamma$ between 7 to 35 days; the average duration of immunity $1/\kappa$ between 1 to 100 years; and the relative strength of immune boosting $\nu$ between 0 and 5. Due to limited quantitative data on immune boosting, we allow immune boosting to be inhibited ($\nu < 1$) or enhanced ($\nu > 1$) during re-exposure to the pathogen. However, we particularly focus on the system dynamics when the strength of immune boosting is relatively weak ($\nu \leq 1$), as the dose of antigens required to stimulate an immune boost is unclear and warrants further examination. We chose the transmission coefficient $\beta$ to range between 200 to 350, encompassing a basic reproductive ratio $R_0 = \beta/(\gamma + \mu)$ between 5.7 to 31.8.

We generated 20,000 parameter sets using LHS. Each parameter set comprised one endemic equilibrium. The procedure used to determine the presence of periodic solutions for each equilibrium follows. We derived analytic expressions for the endemic equilibrium of the SIRWS model. By evaluating the Jacobian of the SIRWS model at the endemic equilibrium, the characteristic equation can be determined. It follows that if the Routh--Hurwitz criteria \citep{Gantmacher1959} were satisfied, all eigenvalues have negative real parts, and the endemic equilibrium is locally asymptotically stable. Otherwise, it is unstable. For analytic expressions of the endemic equilibrium and details on the procedure used to determine stability, the reader is referred to \ref{a:SIRWS-equilibrium}.

The equations of the two-pathogen model were solved using the numerical software XPPAUT \citep{Ermentrout2002} with an adaptive step size Runge--Kutta integrator (Qualst.RK4). Unless specified otherwise, the default model parameters for simulations are detailed in Table \ref{table:params}, similar to the ones used by \cite{Lavine2011} in their study of pertussis. The initial conditions were $X_{SS} = 0.99$, $X_{IS} = X_{SI} = 0.005$, and all other states were set to 0. To focus on the effect of cross-immunity, throughout this paper, we impose the two pathogens to be characteristically indistinguishable ($\beta_1 = \beta_2 = \beta$, $\gamma_1 = \gamma_2 = \gamma$, $\kappa_1 = \kappa_2 = \kappa$, $\nu_1 = \nu_2 = \nu$).

As the two-pathogen model is high-dimensional, our analysis is primarily numerical. Bifurcation analysis was done in XPPAUT with the first 400 years of integration time discarded as a transient with post-processing in \cite{MATLAB2014}. To illustrate different qualitative behaviours in the bifurcation diagrams, we display the variable $X_{IS}$.

\section{Results} \label{sec:results}

\subsection{Absence of sustained oscillations in the single-pathogen SIRWS model for $\nu \leq 1$} \label{results:single_pathogen}

Previously, \cite{Dafilis2012} showed that the SIRWS model described by Equations \eqref{sirws} may exhibit limit cycles for a subset of the parameter space as the birth rate and strength of boosting change simultaneously. Here we expand on this work and vary all parameters simultaneously to determine the presence of periodic solutions throughout the entire parameter space.

\begin{figure}
	\centering
	\includegraphics[width=0.9\textwidth]{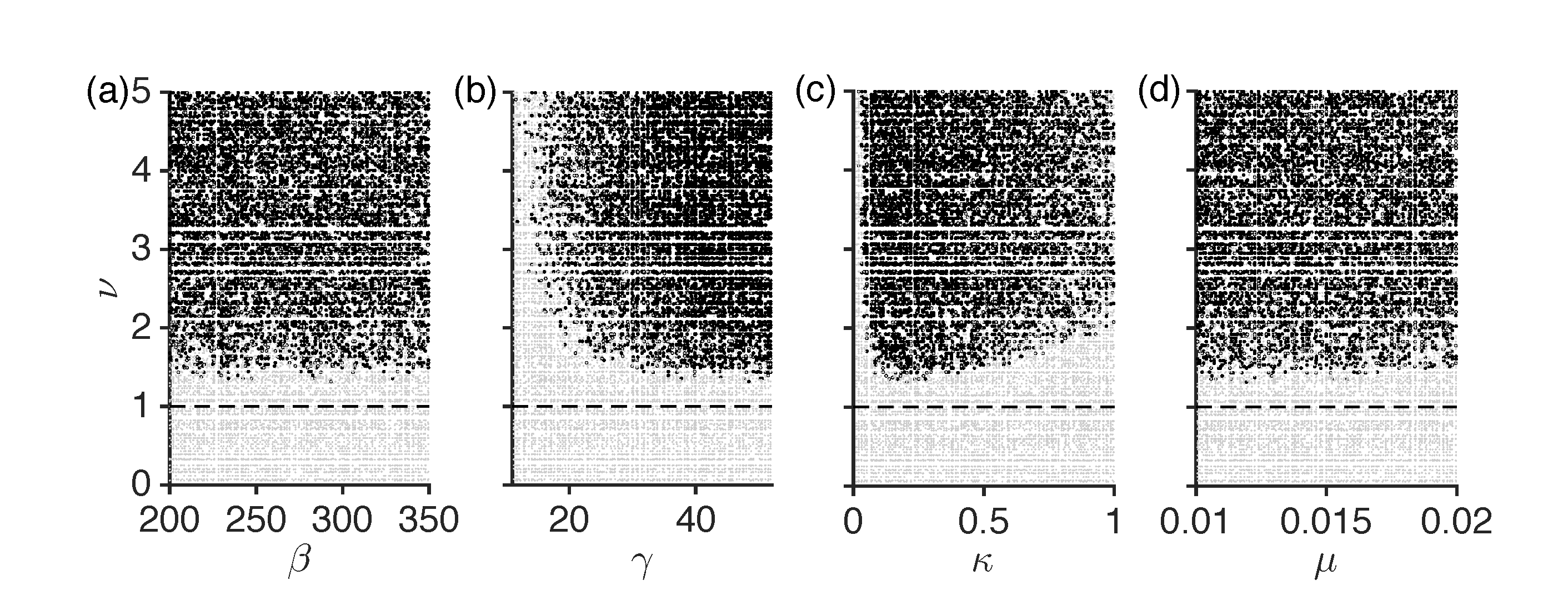} 
	\caption{Numerical calculations run with 20,000 LHS-sampled parameter sets. For each parameter set, the Routh--Hurwitz criterion was employed. A gray dot represents the case when all characteristic roots were found to have negative real parts, and a black dot otherwise. The dashed (- - -) line represents $\nu = 1$.}
	\label{fig:LHS}
\end{figure}

The values of each LHS-generated parameter set are displayed as dots in Figure \ref{fig:LHS}. The colour of the dot represents the stability of the equilibrium consisting of the parameter set: gray when locally asymptotically stable, and black when unstable. We find through numerical calculations that the endemic equilibrium loses stability through a Hopf bifurcation to give rise to sustained oscillations (explored in detail in Section \ref{sec:crossimmunity}). Interestingly, sustained oscillations are observed only when $\nu > 1$, independent of the other parameter values. Thus, the equations of the SIRWS system may only produce periodic solutions when immune boosting is more easily triggered than a naive infection ($\nu \beta I > \beta I$). However, if $\nu \leq 1$, the dynamics of the model are always characterised by a point attractor reached via damped oscillations, thus failing to capture the cyclic dynamics of some infectious diseases, such as pertussis \citep{Lautrop1971}. 

\subsection{Cross-immunity allows for sustained oscillations for $\nu \leq 1$} \label{results:nu_less_than_1}
\label{sec:crossimmunity}

In the absence of cross-immunity ($\sigma_1 = \sigma_2 = 0$) and with symmetric initial conditions, $X_{mn} = X_{nm}$ for all time. By using the following substitutions: 
\begin{align*}
	X_{S \bullet} &= X_{SS} + X_{SI} + X_{SR} + X_{SW} \, , \\
	X_{I \bullet} &= X_{IS} + X_{II} + X_{IR} + X_{IW} \, , \\
	X_{R \bullet} &= X_{RS} + X_{RI} + X_{RR} + X_{RW} \, , \\
	X_{W \bullet} &= X_{WS} + X_{WI} + X_{WR} + X_{WW} \, ,
\end{align*}
the two-pathogen model collapses to the single-pathogen SIRWS model; hence, the presence of periodic solutions for this special case is described in Section \ref{results:single_pathogen}.

\begin{figure}
	\includegraphics[width=\textwidth]{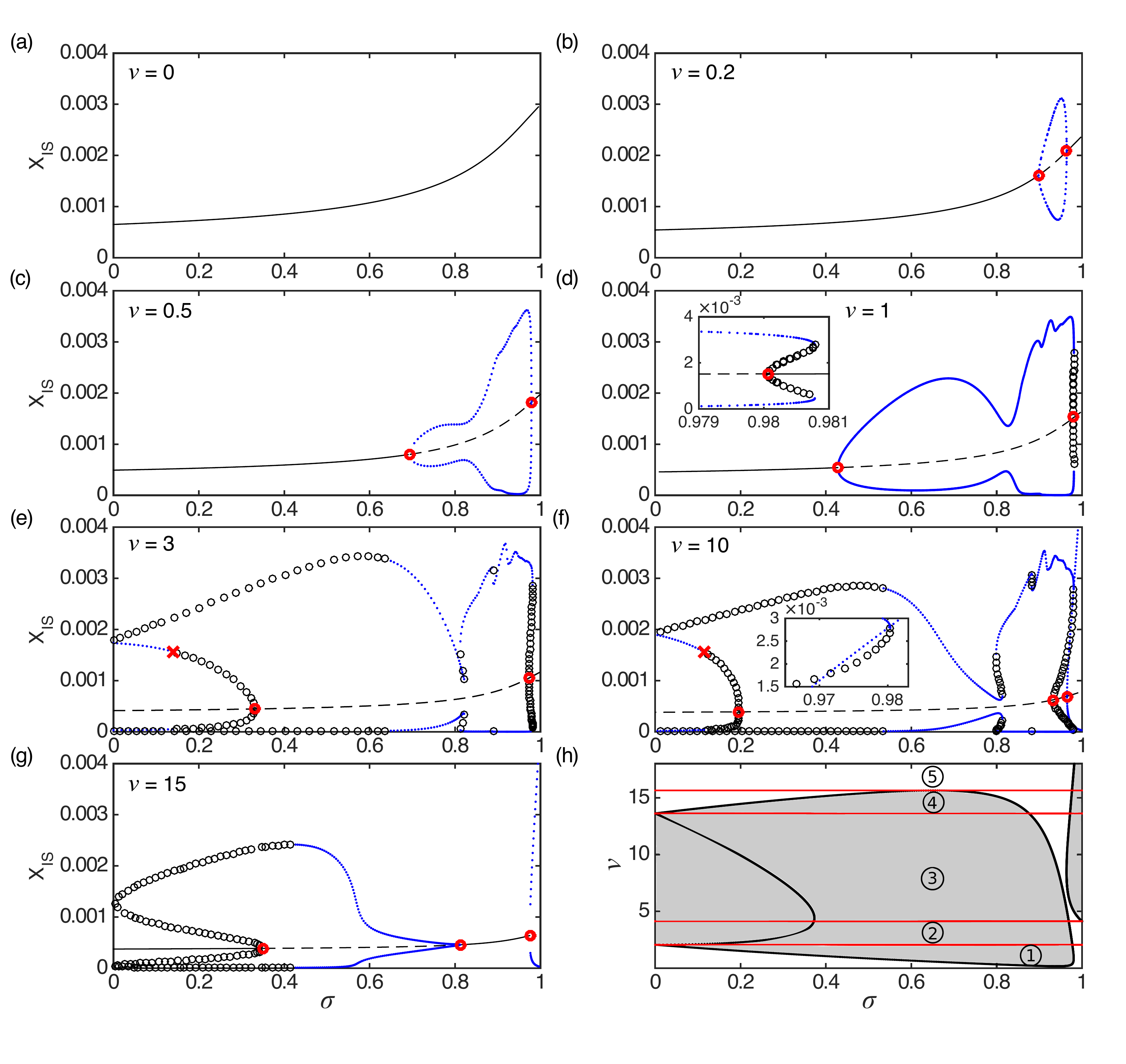}
	\caption{(Colour online) One-parameter bifurcation diagrams of $X_{IS}$ as a function of $\sigma_1 = \sigma_2 = \sigma$ for (a) $\nu = 0$, (b) $\nu = 0.2$, (c) $\nu = 0.5$, (d) $\nu = 1$, (e) $\nu = 3$, (f) $\nu = 10$, and (g) $\nu = 15$. The fixed point is represented by a line: stable (solid) or unstable (dashed). Blue dots indicate stable periodic solutions, and open circles show when periodic solutions are unstable. Red open circles represent a Hopf bifurcation, and a red cross in (e) and (f) represents a torus bifurcation. The loci of Hopf bifurcation points are displayed in (h). Five regions, separated by red lines, are labelled to guide discussion in the main text. ($\mu = 1/80$, $\gamma = 17$, $\kappa = 0.1$, $\beta = 260$, hence $R_0 \approx 15$.)}
	\label{fig:bifurcation}
\end{figure}

\newpage

We now consider the presence of symmetric cross-immunity ($\sigma_1 = \sigma_2 = \sigma > 0$). Figure \ref{fig:bifurcation}(a)--(g) present one-dimensional bifurcation diagrams of $X_{IS}$ as a function of $\sigma$ for seven different values of the boosting parameter $\nu$. In (h), when $\sigma$ and $\nu$ are co-varied, lines of Hopf points describe regions in the parameter space with different dynamical behaviours. The shaded areas indicate where periodic solutions occur. Five regions are labelled based on different qualitative behaviours arising from the Hopf bifurcations:
\begin{enumerate}
	\item $0 \leq \nu < 2.06$
	\item $2.06 \leq \nu < 4.13$
	\item $4.13 \leq \nu < 13.62$ 
	\item $13.62 \leq \nu < 15.64$ 
	\item $\nu \geq 15.64$. 
\end{enumerate}
The values of $\nu$ in Figure \ref{fig:bifurcation}(d)--(g) were chosen to each represent a plane in regions 1 to 4. Collectively, Figure \ref{fig:bifurcation} displays how the family of periodic solutions appearing at $\nu = 0.2$ (b) transforms as $\nu$ increases, and eventually disappears in region 5. 

Region 1 $(0 \leq \nu < 2.06)$ is of particular interest as its parameter space can be dichotomised into regions where immune boosting is inhibited ($\nu < 1$) or enhanced ($\nu > 1$) relative to the force of infection. In the limiting case $\nu = 0$, the fixed point undergoes no bifurcation and remains stable (Figure \ref{fig:bifurcation}(a)). In contrast, as $\nu$ increases to 0.2 (b), the system undergoes a Hopf bifurcation ($\sigma \approx 0.9$), giving rise to sustained oscillations with amplitudes increasing with $\sigma$. The amplitude drops sharply as the system approaches the second Hopf bifurcation ($\sigma \approx 0.96$). As expected, periodic solutions are enclosed within the two Hopf points. When $\nu$ increases further to 0.5 (c), the amplitude as a function of $\sigma$ appears nonlinear as there is a gradual dip between the two Hopf points. At $\nu = 1$ (d), the nonlinear relationship between the amplitude and $\sigma$ is even more noticeable. There is a rapid decrease in amplitude followed by a steep increase between the two Hopf points. Additionally, the family of periodic solutions extends past the second Hopf point which gives rise to a region of bistability (inset, (d)). Which attractor the system settles to over time in the bi-stable region is determined by initial conditions, as exemplified in \ref{a:attractors}. 

\subsection{Sustained oscillations for $\nu > 1$ and their dynamical properties} \label{results:nu_greater_than_1}

Figure \ref{fig:bifurcation}(e) reveals branches of periodic solutions that can assume complicated shapes, showing an interesting and nontrivial sequence of stable and unstable families of oscillations. At $\sigma = 0$, there is no interaction between the two pathogens, and the system's behaviour is described by the one-pathogen model. As \cite{Dafilis2012} have shown, at $\nu = 3$, the system has stable periodic solutions. We find that these periodic solutions remain stable for small values of $\sigma$, losing stability at $\sigma \approx 0.14$ and extending to $\sigma \approx 0.33$. At $\sigma = 1$, which models full cross-immunity between pathogens, there is a stable fixed point, which loses stability at $\sigma \approx 0.98$, giving rise to stable periodic solutions down to $\sigma \approx 0.65$. (Note, there is a narrow bi-stable region for large $\sigma$.) This family of periodic solutions, having lost stability, extends to $\sigma = 0$.

In addition to the two families of periodic solutions, Figure \ref{fig:bifurcation}(f) at $\nu = 10$ features a third family of periodic solutions that extends to $\sigma = 1$. Interestingly, two families of oscillations are seen overlapping in parameter space (inset, (f)), suggesting that multi-stability of different periodic solutions may be present. The variability in the amplitude of oscillations in the limited interval for high $\sigma$ values is also observed for lower values of $\nu$. At an even higher $\nu$ value of 15 (g), the periodic branch connecting the two Hopf points appears more regular, in the sense that the amplitude of the stable locus monotonically decreases to the second Hopf point with increasing $\sigma$. It is interesting to note that the steep increase in amplitude following the rapid decrease, as observed in (d)--(f), is absent in (g). 

In the epidemiological context, the presence of overlapping attractors---each with different amplitudes as observed in the inset of Figure \ref{fig:bifurcation}(f)---may have considerable impact. Consider a scenario where a small perturbation, perhaps as a result of stochastic effects, shifts the population to a different regime. This could translate to an increased health burden. Further, sudden drops---or conversely, spikes---in incidence, as a result of shifts between different attractors or the high variability in amplitude for high values of $\sigma$, would be anticipated to make prediction of disease burden difficult.

\begin{figure}
	\centering
	\includegraphics[width=0.9\textwidth]{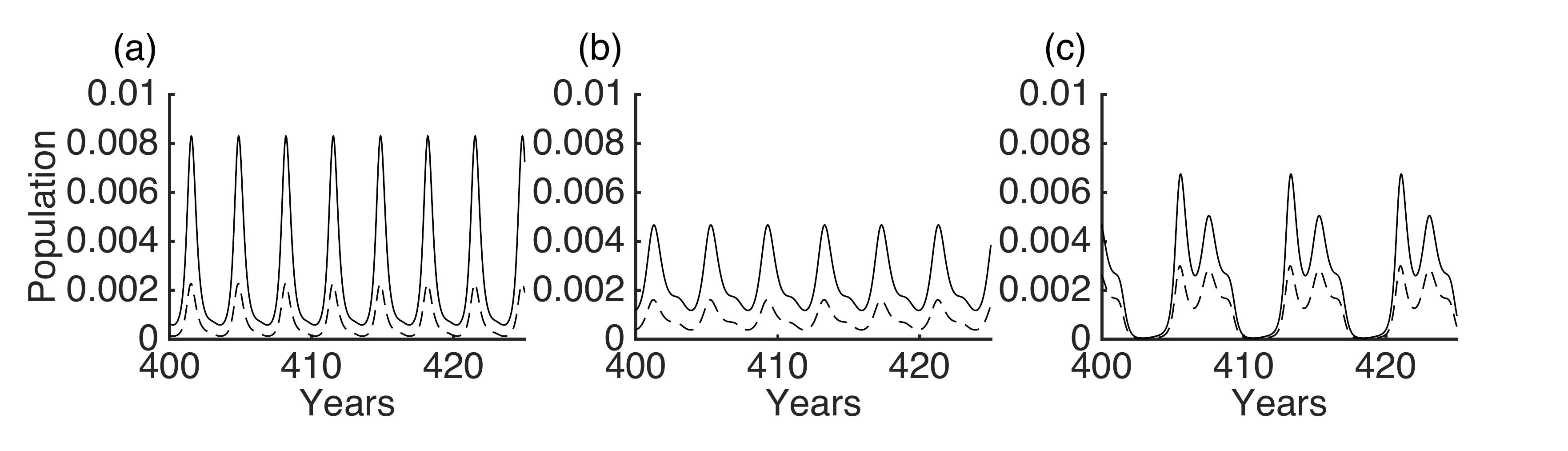} 
	\caption{Time series of the two-pathogen model when $\nu = 1$ and (a) $\sigma = 0.7$, (b) $\sigma = 0.8$, and (c) $\sigma = 0.9$. $X_{I\bullet}$ is shown by a solid line, and $X_{IS}$ in dashed line.}
	\label{fig:timeseries}
\end{figure}

\begin{figure}
	\centering
	\includegraphics[width=0.9\textwidth]{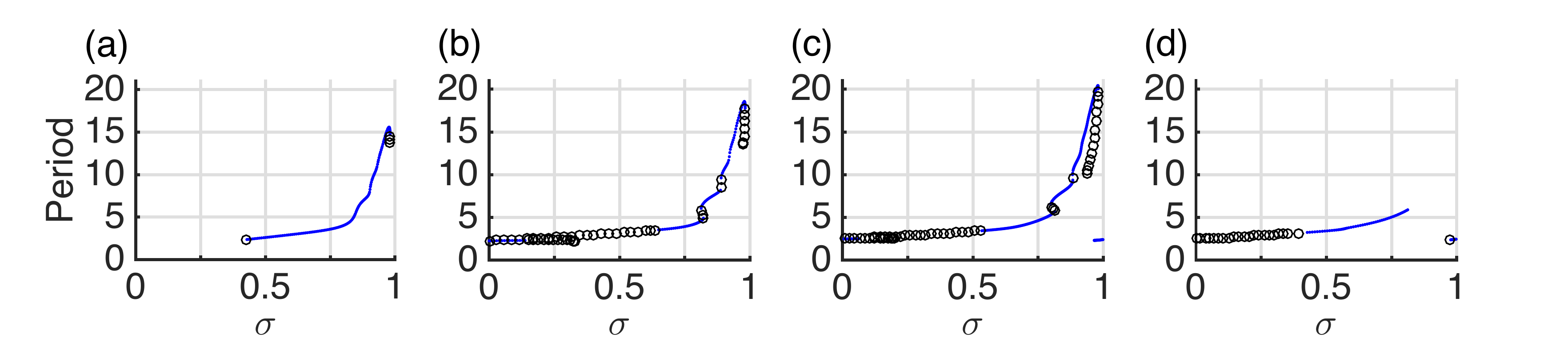} 
	\caption{Diagrams of the period for regions 1, 2, 3, and 4 at (a) $\nu = 1$, (b) $\nu = 3$, (c) $\nu = 10$, and (d) $\nu = 15$, respectively.}
	\label{fig:sigma-period}
\end{figure}

A relatively small change in $\sigma$ can produce strikingly different periodic behaviours in terms of period, amplitude, and wave shape (Figure \ref{fig:timeseries}). The peaks of $X_{I\bullet}$, those infectious with the first pathogen, is reduced by half as $\sigma$ changes from 0.7 to 0.8, and the interepidemic interval is longer, as illustrated in Figure \ref{fig:timeseries}(a)--(b). Comparing (b) to (c), as $\sigma$ increases further to 0.9, the interepidemic interval continues to grow, and the peaks of $X_{IS}$ are higher. The number of peaks in a full cycle also changes from one to two. 

The period of the cycles generally increases with $\sigma$ (Figure \ref{fig:sigma-period}), with the rate of increase steeper for higher values of $\sigma$. Noticeably, the steep increase is missing for $\nu = 15$ (d). Its period appears consistent with the periods of (a)--(c) for lower values of $\sigma$.

\subsection{Dependence on initial conditions} \label{results:IC}

Symmetric initial conditions lead the two pathogens to coexist with the same incidence, period, and amplitude either endemically or while oscillating in phase. However, a small deviation from the symmetric initial conditions can lead to antiphase oscillations with different incidence peaks and amplitudes for the two pathogens---a dramatic change in the dynamical behaviour of the system. 

\begin{figure}
	\centering
	\includegraphics[width=0.9\textwidth]{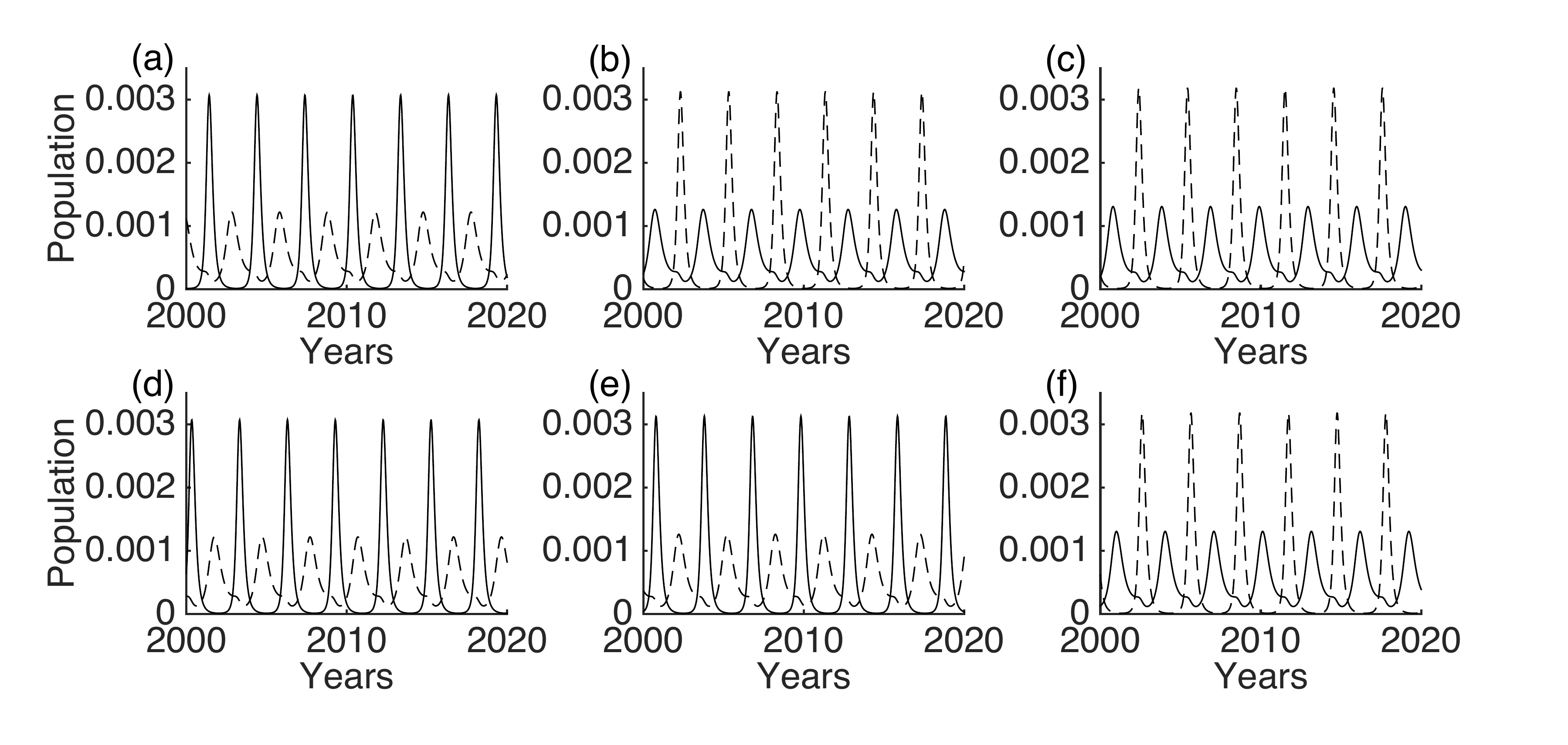} 
	\caption{Scenarios with different $\sigma$ and initial conditions: $X_{SS} = 0.058$, $X_{IS} = X_{SI} = 0.001$, $X_{WS} = X_{SW} = 0.1$, and unspecified states set to 0 ($\nu = 2$). The upper row has ($X_{RS}, X_{SR}$) = (0.41, 0.33), and (a) $\sigma = 0.49$, (b) $\sigma = 0.50$, and (c) $\sigma = 0.51$. The lower row has ($X_{RS}, X_{SR}$) = (0.405, 0.335) and (d) $\sigma = 0.49$, (e) $\sigma = 0.50$, and (f) $\sigma = 0.51$.}
	\label{fig:IC}
\end{figure}

An example is shown in Figure \ref{fig:IC}, where simulations from the top row, (a)--(c), are run with the same set of initial conditions, and the bottom row (d)--(f) with a different set. In a similar fashion, simulations paired by each column ((a) and (d); (b) and (e); and (c) and (f)) are run with the same value of $\sigma$. The oscillations with smaller amplitude also coincide with a lower incidence peak. Small changes in initial conditions (comparing (b) and (e)) or changes in parameterisation (comparing (a) and (b)) can result in different dominant pathogens. The former suggests that multi-stability may be present.

This may manifest in an epidemiological context when considering a vaccination campaign, in which susceptibles are impulsively shifted to resistant classes. Subtle differences in proportions of the population that are shifted may lead to dramatic changes in the dominance of one pathogen over another.

\subsection{Dynamical behaviour when cross-immunity is not symmetric}

\begin{figure}
	\centering
	\includegraphics[width=0.9\textwidth]{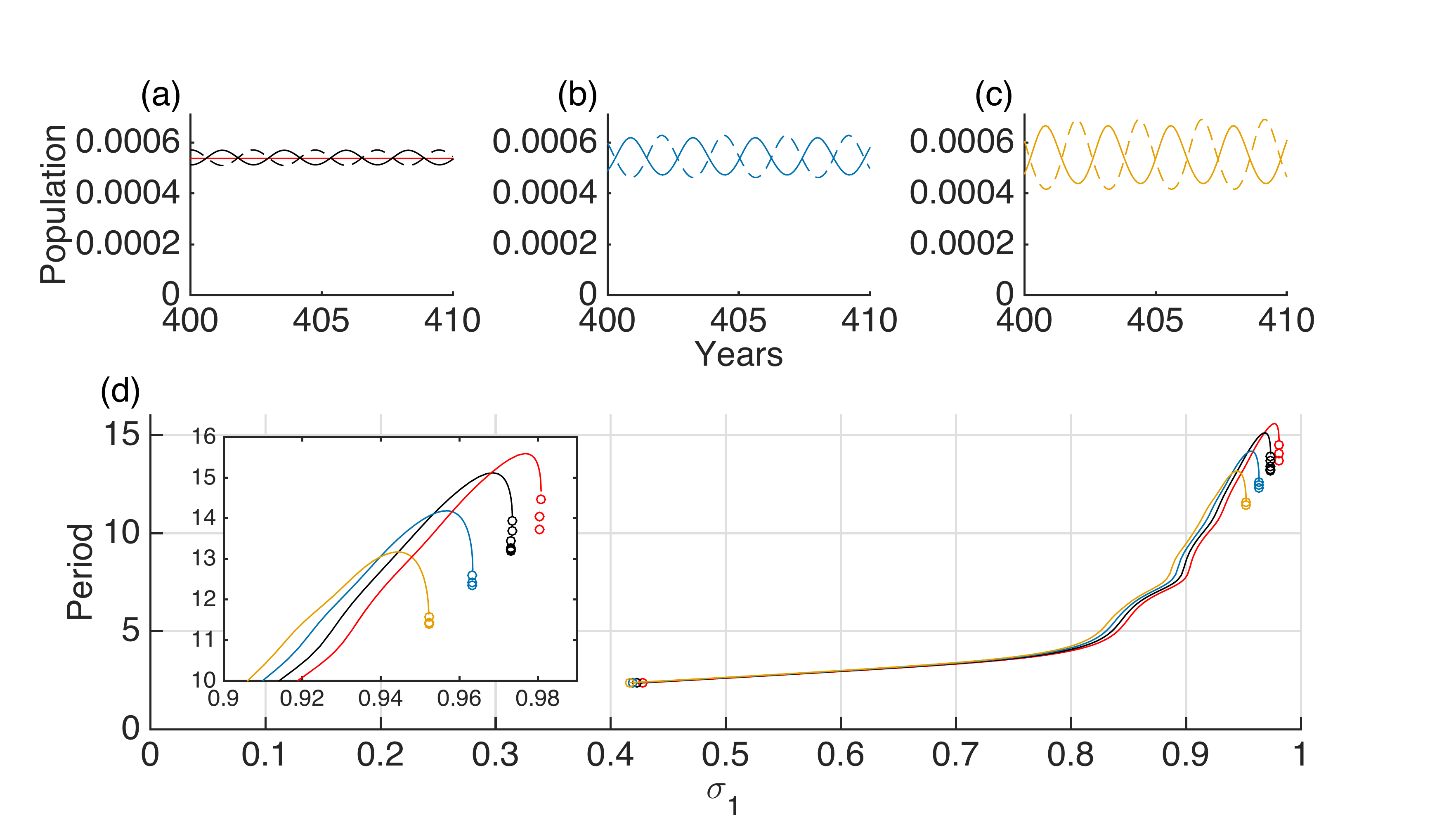} 
	\caption{(Colour online) Time series of the two-pathogen model with asymmetric cross-immunity when $\nu = 1$, $\sigma_1 = 0.42$, and $\sigma_2 = \sigma_1 + \epsilon$. $X_{IS}$ (solid) and $X_{SI}$ (dashed) are plotted for (a) $\epsilon = 0$ (red) and 0.01 (black), (b) $\epsilon = 0.02$ (blue), and (c) $\epsilon = 0.03$ (orange). For varying $\sigma_1$, the corresponding periods of the sustained oscillations for $X_{IS}$ are shown in (d), represented in the same colours.}
	\label{fig:epsilon}
\end{figure}

Here we briefly study the effect of two antigenically distinct pathogens. We model asymmetry in cross-immunity by allowing $\sigma_1$ and $\sigma_2$ to differ. We write $\sigma_2 - \sigma_1 = \epsilon$, noting that $\epsilon = 0$ corresponds to the earlier sections where the two pathogens are antigenically indistinguishable. We start by enforcing $\sigma_1 = 0.42$, a value in the neighbourhood of the Hopf point ($\sigma_H = 0.428$; Figure \ref{fig:bifurcation}(d)) when cross-immunity is symmetric. We set $\sigma_2 = \sigma_1 + \epsilon$, for $\epsilon \in \{0.01, 0.02, 0.03\}$. As $\epsilon$ grows, the two pathogens can exhibit antiphase oscillations with different amplitudes, as illustrated in Figure \ref{fig:epsilon}(a)--(c). In our example, the pathogen conferring more cross-immunity (the second pathogen) oscillates at higher amplitude. The difference in amplitudes increases with $\epsilon$. This indicates that in addition to changes in initial conditions as discussed in Section \ref{results:IC}, antigenic asymmetry may play a role in determining the dominant pathogen. Furthermore, note that both pathogens display oscillatory behaviour even though $\sigma_1 < \sigma_H$, indicating that asymmetry can change the conditions under which a pathogen can sustain oscillations.

Allowing $\sigma_1$ to vary, the periods of oscillations for each value of $\epsilon$ are shown in Figure \ref{fig:epsilon}(d) as a function of $\sigma_1$. For intermediate values of $\sigma_1$, different values of $\epsilon$ produce negligible differences in the periods of oscillations, as shown by the nearly overlapping lines for $0.4 \lesssim \sigma_1 \lesssim 0.8$. However, for large values of $\sigma_1$, a small change in $\epsilon$ can result in a considerable difference in period, as observed in the inset. Moreover, two distinct values of $\sigma_1$ can sustain the same period. For example, when $\epsilon = 0.03$ (orange curve), a period of 12 years can be sustained at $\sigma_1 \approx 0.927$ and $\sigma_1 \approx 0.952$.

\section{Discussion} \label{sec:discussion}

We have demonstrated that a two-pathogen model of disease transmission with immune boosting and cross-immunity can produce behaviours that are qualitatively similar to the incidence patterns of \textit{B.~pertussis} and \textit{B.~parapertussis} as observed by \cite{Lautrop1971}, including antiphase oscillatory dynamics. Our analysis shows that cross-immunity can induce sustained oscillations when immune boosting is both enhanced ($\nu > 1$) or inhibited ($\nu < 1$) by re-exposure. In contrast, sustained oscillations were generated in the single-pathogen SIRWS system only when immune boosting was enhanced, i.e., $\nu > 1$ (See Figure \ref{fig:LHS}, \cite{Dafilis2012}, and \cite{Lavine2011}). Our study finds that the level of reduced susceptibility ($\sigma$) has a significant effect on the shape, amplitude, and period of oscillations. Interestingly, initial conditions can determine which one out of the two characteristically and antigenically similar pathogens is dominant.

Our study underscores the importance of developing a better understanding of immunological interactions between pathogens to better inform model-based interpretations of epidemiological data. We have shown that the degree of cross-immunity and strength of immune boosting may determine the epidemiological behaviour of a disease with both endemic and cyclic (oscillatory) dynamics supported. Under weak levels of immune boosting, a high degree of cross-immunity may introduce a cyclic regime (Figure \ref{fig:bifurcation}). In contrast, as the rate of immune boosting grows to be equal to the force of infection, only intermediate levels of cross-immunity may be required for cyclic behaviour. Determining the likely strength of immune boosting would provide guidance on its role in the maintenance of immunity.

We have not investigated in detail the basins of attraction for the different dynamical regimes identified in our study. Bifurcation analysis of the two-pathogen model performed here indicates that multi-stability is present. The complicated shapes of the Hopf bifurcation branches suggest that the structure of the basins of attraction may be similarly intricate. Their structure may inform the system's sensitivity to initial conditions.

The mechanisms behind immunity are not well understood. Different types of immunity, such as natural and vaccine-induced, act together to maintain population immunity, which is further complicated by immune boosting. Incorporation of vaccination in future work may untangle effects arising from the interplay between natural and vaccine-induced immunity. As multiple factors are capable of generating sustained oscillations in dynamic transmission models, how much immune boosting, cross-immunity, and vaccination contribute to the overall observed recurrence of epidemics of infectious diseases remains an open question.

\section*{Acknowledgements}
We thank Patricia Campbell from the University of Melbourne for insightful discussions. Tiffany Leung is supported by a Melbourne International Research Scholarship from the University of Melbourne and has received funding from a National Health and Medical Research Council funded Centre for Research Excellence in Infectious Diseases Modelling to Inform Public Health Policy (1078068). James M.~McCaw is supported by an Australian Research Council Future Fellowship (110100250).

\newpage
\appendix
\section{The two-pathogen model} \label{a:two-SIRWS}

In this appendix, the equations for the two-pathogen SIRWS model are given. The equations track the immune history of each class in a population, given by
\begin{subequations}
\label{eq:two-SIRWS}
\begin{align}
	\frac{dX_{SS}}{dt} &= \mu + 2\kappa_1 X_{WS} + 2\kappa_2 X_{SW} - (\lambda_1 + \lambda_2 + \mu) X_{SS}  \, , \\
	\frac{dX_{IS}}{dt} &= \lambda_1 X_{SS} + 2\kappa_2 X_{IW} - ((1 - \sigma_2) \lambda_2 + \gamma_1 + \mu) X_{IS} \, , \\
	\frac{dX_{RS}}{dt} &= \gamma_1 X_{IS} + 2\kappa_2 X_{RW} + \nu_1 \lambda_1 X_{WS} - ((1 - \sigma_2) \lambda_2 + 2\kappa_1 + \mu) X_{RS} \, , \\
	\frac{dX_{WS}}{dt} &= 2\kappa_1 X_{RS} + 2\kappa_2 X_{WW} - ((1 - \sigma_2) \lambda_2 + \nu_1 \lambda_1 + 2\kappa_1 + \mu) X_{WS} \, , \\
	\frac{dX_{SI}}{dt} &= \lambda_2 X_{SS} + 2\kappa_1 X_{WI} - ((1 - \sigma_1) \lambda_1 + \gamma_2 + \mu) X_{SI} \, , \\
	\frac{dX_{II}}{dt} &= (1 - \sigma_1) \lambda_1 X_{SI} + (1 - \sigma_2) \lambda_2 X_{IS} - (\gamma_1 + \gamma_2 + \mu) X_{II} \, , \\
	\frac{dX_{RI}}{dt} &= \gamma_1 X_{II} + (1 - \sigma_2) \lambda_2 X_{RS} + \nu_1 \lambda_1 X_{WI} - (2\kappa_1 + \gamma_2 + \mu) X_{RI} \, , \\
	\frac{dX_{WI}}{dt} &= 2\kappa_1 X_{RI} + (1 - \sigma_2) \lambda_2 X_{WS} - (\nu \lambda_1 + 2\kappa_1 + \gamma_2 + \mu) X_{WI} \, , \\
	\frac{dX_{SR}}{dt} &= \gamma_2 X_{SI} + 2\kappa_1 X_{WR} + \nu_2 \lambda_2 X_{SW} - ((1 - \sigma_1) \lambda_1 + 2\kappa_2 + \mu) X_{SR} \, , \\
	\frac{dX_{IR}}{dt} &= \gamma_2 X_{II} + (1 - \sigma_1) \lambda_1 X_{SR} + \nu_2 \lambda_2 X_{IW} - (2\kappa_2 + \gamma_1 + \mu) X_{IR} \, , \\
	\frac{dX_{RR}}{dt} &= \gamma_1 X_{IR} + \gamma_2 X_{RI} + \nu_1 \lambda_1 X_{WR} + \nu_2 \lambda_2 X_{RW} - (2\kappa_1 + 2\kappa_2 + \mu) X_{RR} \, , \\
	\frac{dX_{WR}}{dt} &= \gamma_2 X_{WI} + 2\kappa_1 X_{RR} + \nu_2 \lambda_2 X_{WW} - (\nu_1 \lambda_1 + 2\kappa_1 + 2\kappa_2 + \mu) X_{WR} \, , \\
	\frac{dX_{SW}}{dt} &= 2\kappa_1 X_{WW} + 2\kappa_2 X_{SR} - ((1 - \sigma_1)\lambda_1 + \nu_2 \lambda_2 + 2\kappa_2 + \mu) X_{SW} \, , \\
	\frac{dX_{IW}}{dt} &= 2\kappa_2 X_{IR} + (1 - \sigma_1)\lambda_1 X_{SW} - (2\kappa_2 + \nu_2 \lambda_2 + \gamma_1 + \mu) X_{IW} \, , \\
	\frac{dX_{RW}}{dt} &= \gamma_1 X_{IW} + 2\kappa_2 X_{RR} + \nu_1 \lambda_1 X_{WW} - (\nu_2 \lambda_2 + 2\kappa_1 + 2\kappa_2 + \mu) X_{RW} \, , \\
	\frac{dX_{WW}}{dt} &= 2\kappa_1 X_{RW} + 2\kappa_2 X_{WR} - (\nu_1 \lambda_1 + \nu_2 \lambda_2 + 2\kappa_1 + 2\kappa_2 + \mu) X_{WW} \, ,
\end{align}
\end{subequations}
where
\begin{subequations}
\begin{align}
	\lambda_1 &= \beta_1(X_{IS} + (1-\alpha_1) (X_{II} + X_{IR} + X_{IW})) \, , \\
	\lambda_2 &= \beta_2(X_{SI} + (1-\alpha_2) (X_{II} + X_{RI} + X_{WI})) \, .
\end{align}
\end{subequations}
To incorporate cross-immunity through a reduction in infectivity, or the ability to transmit the infection, the force of infection $\lambda_i$ becomes the transmission rate times a weighted sum of those infectious with pathogen $i = 1,2$. Those who have experience with infection with pathogen $j$ become $(1 - \alpha_i)$ times less infectious than those with no experience, $0 \leq \alpha_i \leq 1$. Throughout the paper, we impose $\alpha_1 = \alpha_2 = 0$ (no reduced infectivity).

\section{Equilibrium of the SIRWS model} \label{a:SIRWS-equilibrium}

In this section, we present the endemic equilibrium of the SIRWS model and determine its stability using the Routh--Hurwitz criteria \citep{Gantmacher1959}. Expressions of the endemic equilibrium for the SIRWS model are given by
\begin{subequations}
\label{sirws-endemic-eq}	
\begin{align}
	S^* &= \frac{\gamma + \mu}{\beta} \, , \\
	I^* &= \frac{\sqrt{c_0^2 + c_1 \nu + c_2^2 \nu^2} - [4\kappa^2 + (4\kappa + \mu)(\gamma + \mu) - \mu \nu(\beta - \gamma - \mu)]}{2\beta(\gamma + \mu) \nu} \, , \\
	R^* &= \frac{(2\kappa + \mu)(\gamma + \mu) + 2\kappa \gamma}{4\kappa (\gamma + \mu)}(\beta - \gamma - \mu) + \frac{\gamma + \mu + 2\kappa}{4\kappa \beta (\gamma + \mu) \nu} c_0 \left( 1 - \sqrt{1 + c_1 \nu + c_2 \nu^2} \right) \, , \\
	W^* &= \frac{\sqrt{c_0^2 + c_1 \nu + c_2^2 \nu^2} - [4\kappa^2 + (4\kappa + \mu)(\gamma + \mu) + \mu \nu(\beta - \gamma - \mu)]}{4\kappa \beta \nu} \, ,
\end{align}
\end{subequations}
where
\begin{align*}
	c_0 &= (2\kappa + \mu)^2 + \gamma (4\kappa + \mu) \, , \\
	c_1 &= \frac{2(\beta - \gamma - \mu)[\mu(2\kappa + \mu)^2 + \gamma(8\kappa^2 + 4\kappa \mu + \mu^2)]}{[(2\kappa + \mu)^2 + \gamma (4\kappa + \mu)]^2} \, , \\
	c_2 &= \frac{[\mu(-\beta + \gamma + \mu)]^2}{[(2\kappa + \mu)^2 + \gamma (4\kappa + \mu)]^2} \, .
\end{align*}

The Jacobian of the SIRWS system is
\begin{align*}
	\mathbf{J} = \left( \begin{array}{cccc}
	-(\mu + \beta I) & -\beta S & 0 & 2\kappa \\
	\beta I & -(\gamma + \mu - \beta S) & 0 & 0 \\
	0 & \gamma + \beta \nu W & -(2\kappa + \mu) & \beta \nu I \\
	0 & -\beta \nu W & 2\kappa & -(2\kappa + \mu + \beta \nu I) \end{array} \right) \, .
\end{align*}
By evaluating $\mathbf{J}$ at the endemic equilibrium, the corresponding characteristic equation is given by
\[ \det (\mathbf{J} - \Lambda \mathbf{I}) = 0 \, , \]
where $\mathbf{I}$ is the identity matrix, and $\Lambda$ is the eigenvalue. Notice that $(\mu + \Lambda)$ can be factored out from the first row of $(\mathbf{J} - \Lambda \mathbf{I})$ by the row addition $\textrm{R}_1 + \textrm{R}_2 + \textrm{R}_3 + \textrm{R}_4 \rightarrow \textrm{R}_1$, giving one negative eigenvalue $\Lambda_1 = -\mu$. The characteristic equation simplifies to
\begin{align} \label{a:char_eqn}
	(\mu + \Lambda)(\Lambda^3 + a_2 \Lambda^2 + a_1 \Lambda + a_0) = 0\, ,
\end{align}
where
\begin{align*}
	a_2 &= \beta I^* (1 + \nu) + 2(2\kappa + \mu) \, , \\
	a_1 &= \beta I^* [2(2\kappa + \mu) + \gamma + \beta \nu I^* + \mu \nu ] + (2\kappa + \mu)^2 \, , \\
	a_0 &= \beta I^* [ (2\kappa + \mu)^2 + \mu \beta \nu I^* + \gamma(4\kappa + \mu + \beta \nu I^*) + 2\kappa \beta \nu W^*] \, .
\end{align*}
The equilibrium is locally asymptotically stable if the Routh--Hurwitz criteria are satisfied, and unstable otherwise.

\section{Coexisting attractors} \label{a:attractors}

\begin{figure}[h]
	\centering
	\includegraphics[width=0.66\textwidth]{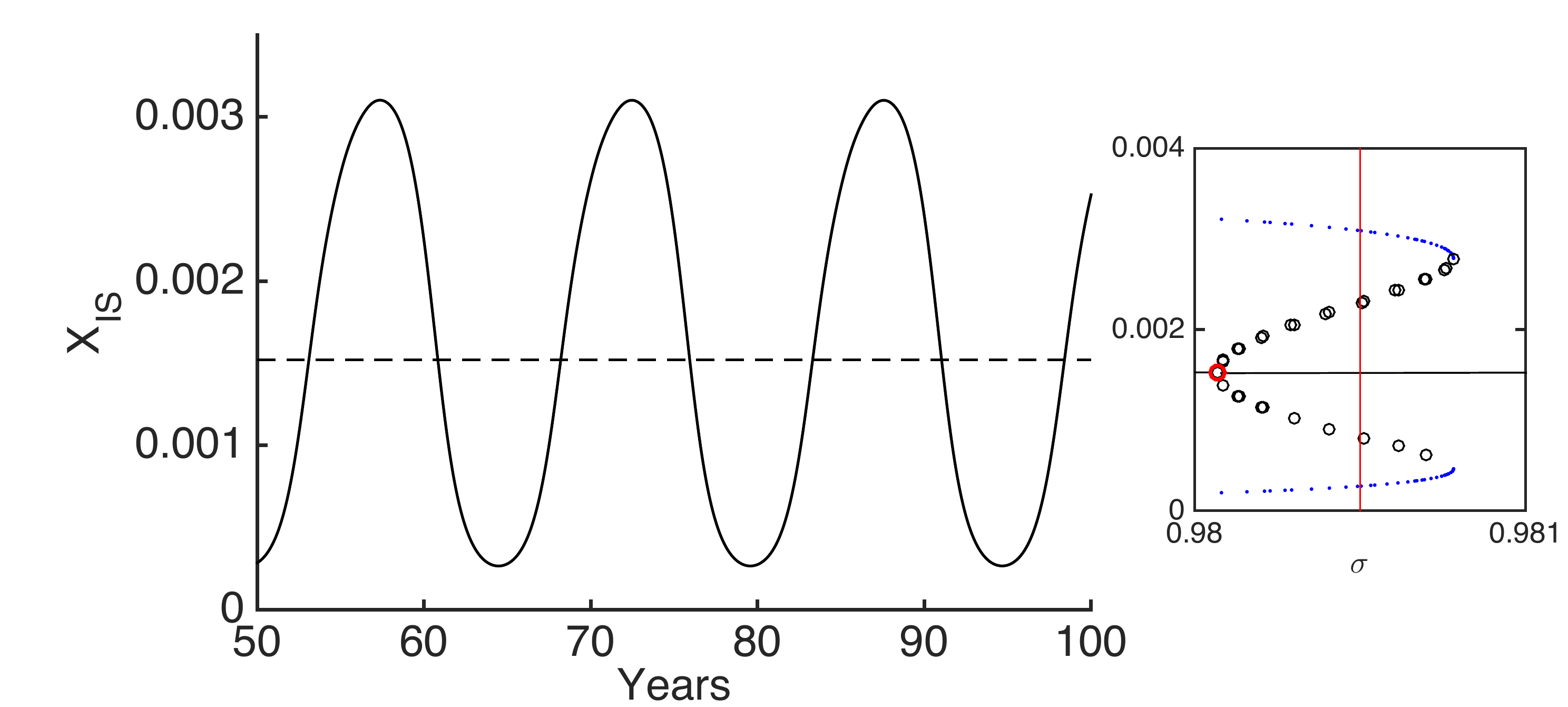}
	\caption{(Colour online) Periodic solutions (solid) with initial conditions $X_{\textrm{periodic}}$, and point attractor (dashed) solutions with initial conditions $X_{\textrm{point}}$. The red line in the bifurcation diagram (inset) shows the value of $\sigma = 0.9805$ at $\nu = 1$.}
	\label{a:bistable}
\end{figure}

This section provides an example of bi-stability in the two-pathogen model (Figure \ref{a:bistable}) with $\nu = 1$, $\sigma_1 = \sigma_2 = \sigma = 0.9805$, and initial population size
\begin{multline*}
	X = (X_{SS}\, , X_{IS}\, , X_{RS}\, , X_{WS}\, , 
		X_{SI}\, , X_{II}\, , X_{RI}\, , X_{WI}\, , \\
		X_{SR}\, , X_{IR}\, , X_{RR}\, , X_{WR}\, , 
		X_{SW}\, , X_{IW}\, , X_{RW}\, , X_{WW}) \, .
\end{multline*}
Initial conditions used to generate periodic solutions are 
\begin{multline*}
X_{\textrm{periodic}} = (0.0571, 0.0019, 0.4443, 0.0944, 0.0014, 0.0000, 0.0002, 0.0000, \\
0.2163, 0.0001, 0.0367, 0.0079, 0.1156, 0.0001, 0.0198, 0.0043).
\end{multline*}
Initial conditions for a point attractor are 
\begin{multline*}
X_{\textrm{point}} = (0.0568, 0.0015, 0.3329, 0.0996, 0.0015, 0.0000, 0.0002, 0.0001, \\
0.3329, 0.0002, 0.0441, 0.0133, 0.0996, 0.0001, 0.0133, 0.0040)
\end{multline*}

\section*{References}

\bibliographystyle{elsarticle-harv} 
\bibliography{library}

\end{document}